\pdfoutput=1
\documentclass[a4paper]{rsproca_new}%%%%where rsproca is the template name

\usepackage{amsthm}
\usepackage{graphicx}
\usepackage[title]{appendix}
\usepackage{mathtools}
\usepackage{siunitx}
\usepackage{upgreek}
\usepackage{xcolor}
\usepackage{cancel}
\usepackage{bbm}
\usepackage{float}

\usepackage[T1]{fontenc}
\usepackage[utf8]{inputenc}

\def\1{{\bf 1}}\def\0{{\bf 0}}
\def\p{{\partial}}

\titlehead{Research}

\begin{document}

%%%% Article title to be placed here
\title{Rotational mobility in spherical membranes: The interplay between Saffman-Delbr{\"u}ck length and inclusion size}

\author{%%%% Author details
Marco Vona and Eric Lauga}

%%%%%%%%% Insert author address here
\address{Department of Applied Mathematics and Theoretical Physics, University of Cambridge, Cambridge, UK}

%%%% Subject entries to be placed here %%%%
\subject{mathematical physics, fluid dynamics}

%%%% Keyword entries to be placed here %%%%

%%%% Insert corresponding author and its email address}
\corres{Eric Lauga\\
\email{e.lauga@damtp.cam.ac.uk}}
%%%% Abstract text to be placed here %%%%%%%%%%%%
\begin{abstract}
The  {mobility} of  {particles in} fluid membranes is a fundamental aspect of many biological and physical processes.  In a 1975 paper~\cite{SaffmanFlow}, Saffman and Delbrück demonstrated how the presence of external Stokesian solvents is crucial in regularising the apparently singular flow within an infinite flat membrane. In {the present} paper, we extend this classical work and compute the rotational mobility of a rigid  {finite-sized particle} located inside a spherical membrane embedded in Stokesian solvents.  Treating the particle as {a spherical cap}, we solve for the flow semi-analytically as a function of  {the} Saffman-Delbr{\"u}ck (SD) length (ratio of membrane to solvent viscosity)  {and the solid angle formed by the particle.} We study the dependence of the mobility and flow on inclusion size and SD length, recovering the flat-space {mobility} as a special case. Our results will be applicable to a range of biological problems including rotational Brownian motion, the dynamics of lipid rafts, and the motion of aquaporin channels in response to water flow. Our method will provide a novel way of measuring a membrane’s viscosity from the rotational diffusion of large inclusions, for which the commonly used planar Saffman-Delbrück theory does not apply.
\end{abstract}
\maketitle

\section{Introduction}\label{sec:Intro}
\begin{figure}[t]
\includegraphics[width = 1\textwidth]{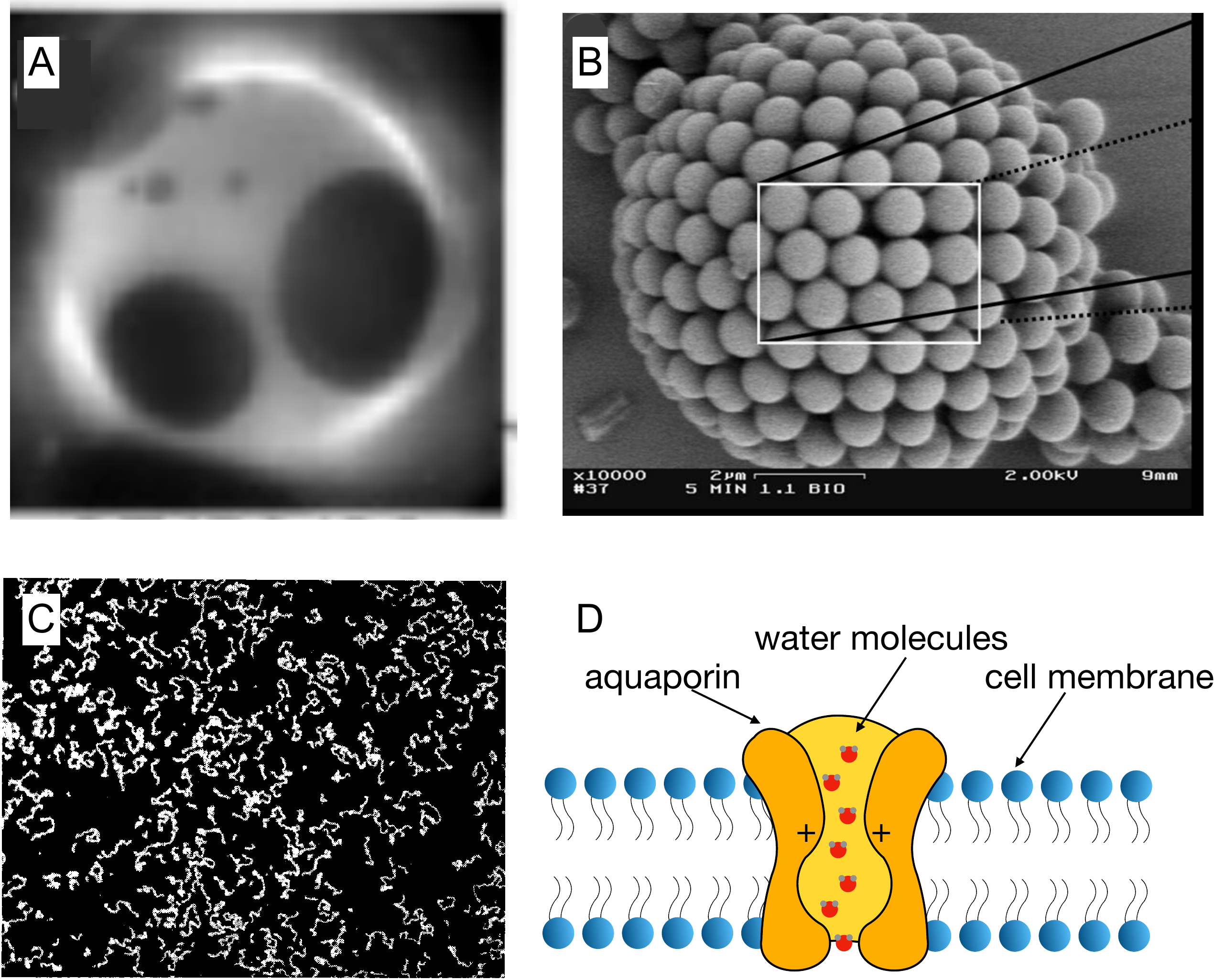}
\caption{{{Diverse experimental examples of membrane inclusions}: (A) {Liquid domains ($0.3-10\ \mu$m in radius) in giant unilamellar vesicles ($15-50\ \mu$m in radius),  with darker regions corresponding to higher viscosity~\cite{cicuta2007diffusion}} (B) Colloids adsorbed on the fluid-fluid interfaces of emulsion droplets may be locked together to form a selectively permeable capsule~\cite{dinsmore2002colloidosomes}. (C) Semidilute solution of  {DNA} electrostatically bound to a cationic lipid membrane and diffusing in-plane~\cite{maier2000dna}.  (D) {Schematic representation of human aquaporin,}  {which facilitates} efficient and specific passive permeation of water and other small uncharged solutes across the cell membrane. {In this paper, we concern ourselves with rigid inclusions only}.
}
}

\label{Fig:Experimental Situations Collage}
\end{figure}

The  {mobility} of  {macroscopic {inclusions}} located inside fluid membranes plays a role in many physical and biological processes, {such as} the kinetics of  {liquid} domains in giant unilamellar vesicles~\cite{veatch2005miscibility, veatch2005seeing} (Fig.~\ref{Fig:Experimental Situations Collage}A), the formation of finite-sized compartments in surface monolayers consisting of multiple chemical components~\cite{stone1998hydrodynamics, mcconnell1991structures}, the kinetics of colloids {adsorbed} on liquid droplets~\cite{dinsmore2002colloidosomes} (Fig.~\ref{Fig:Experimental Situations Collage}B), the rotational diffusion of  membrane-bound polymers~\cite{maier2000dna} (Fig.~\ref{Fig:Experimental Situations Collage}C) and  particles~\cite{spooner2000rotational},  {and the postulated  motion of aquaporin channels~\cite{OptimizingPermeability, WaterPermeation, AquaporinsStructure} in response to water flow~\cite{WaterPermeation}  (Fig.~\ref{Fig:Experimental Situations Collage}D).}  {Biological membranes often  display curvature~\cite{rahimi2013curved}, which may be either  {intrinsic}~\cite{MembraneBook} or  {the} result of stochastic fluctuations, as seen for example in the ``flicker phenomenon'' of erythrocytes~\cite{brochard1975frequency}}. The prediction of particle  {mobility} therefore requires  {formulating} a hydrodynamic theory for flows inside curved membranes.

In the absence of shear orthogonal to the  {membrane}~\cite{SaffmanFlow, MembraneBook}, as in the case  {of} lipid {bilayers}~\cite{HelfrichEnergy, arroyo2009relaxation}, it is appropriate to model membranes as two-dimensional fluids subject to internal   viscous stresses and embedded within three-dimensional fluids referred to as {solvents}~\cite{SaffmanFlow, LipidBilayer, CurvedNSDerivation, powers2010dynamics, MembraneIsImpermeableToSolute, MembraneProperties, al2023morphodynamics}.  These  {solvents} are coupled to the membrane by the no-slip and stress-balance boundary conditions~\cite{CurvedNSDerivation, powers2010dynamics} which, unlike for a simple fluid-fluid interface, must also account for the membrane viscous stresses~\cite{HelfrichBendingEnergy, santiago2018stresses, rangamani2013interaction}. {As often done, we  consider the case of membranes that are incompressible and impermeable to the solvents~\cite{MembraneIsImpermeableToSolute, henle2008effect}}.

 The presence of the solvents  is not only biologically relevant,  {but also} offers a resolution to the mathematical issues connected with {translational} mobility in an infinite membrane. Without solvents, a particle embedded in an infinite flat membrane would not have a well-defined {translational} mobility, due to the    Stokes paradox~\cite{FluidsBook}.  {Indeed, the two-dimensional Stokes flow around a body diverges logarithmically when the force on the body is nonzero, resulting in a theoretically infinite mobility}. A resolution to the paradox was offered by Saffman and Delbr{\"u}ck in   a  {classical paper}~\cite{SaffmanFlow}.  {They} showed  {that} the coupling to the viscous  {solvents} below and above the membrane regularises the problem by introducing a natural  {cut-off} length scale  {for the logarithmic divergence}, now known as the Saffman-Delbr{\"u}ck length and given by  $\ell_{\text{SD}}\equiv \eta/\mu$, where  $\eta$  is the two-dimensional membrane  viscosity and $\mu$  {is} the 
 solvent viscosity~\cite{SaffmanDetailedCalculations}.  The features of the flow then strongly depend on the relative magnitude  {of} the Saffman-Delbr{\"u}ck length and the other length scales in the problem (e.g.~local radius of curvature, particle size)~\cite{daniels2007diffusion, henle2008effect}.  {Note that membrane inertia, or the finite size of the membrane, may also  be used to regularise the problem~\cite{SaffmanFlow}}.
 Saffman and Delbr{\"u}ck's discovery sparked a flurry of activity on particle mobility in different biological setups and geometries~\cite{CurvedNSDerivation, CylinderInMembrane}, including spherical~\cite{samanta2021vortex, bagaria2022dynamics} or tubular~\cite{jain2023force} membranes. {Further studies considered slender~\cite{levine2004mobility, shi2024drag} or active~\cite{hosaka2017lateral} inclusions, rigid boundaries inside the ambient fluids~\cite{stone1998hydrodynamics}, the effect of membrane deformability~\cite{arroyo2009relaxation, rahimi2013curved} and elasticity~\cite{brown2008elastic}}. {In all cases}, solutions to the hydrodynamic problem must account for the extreme variability in inclusion size, which may range from that of a single peptide or lipid  {(about $10$ nm in size)} to larger bodies such as protein aggregates~\cite{shi2024drag, shi2022hydrodynamics}  and  {liquid} domains ($0.3-10\ \mu$m in radius)~\cite{veatch2005seeing, 
 cicuta2007diffusion,  komura2014physical}.
 
\begin{figure}[t]
\includegraphics[width = 1\textwidth]{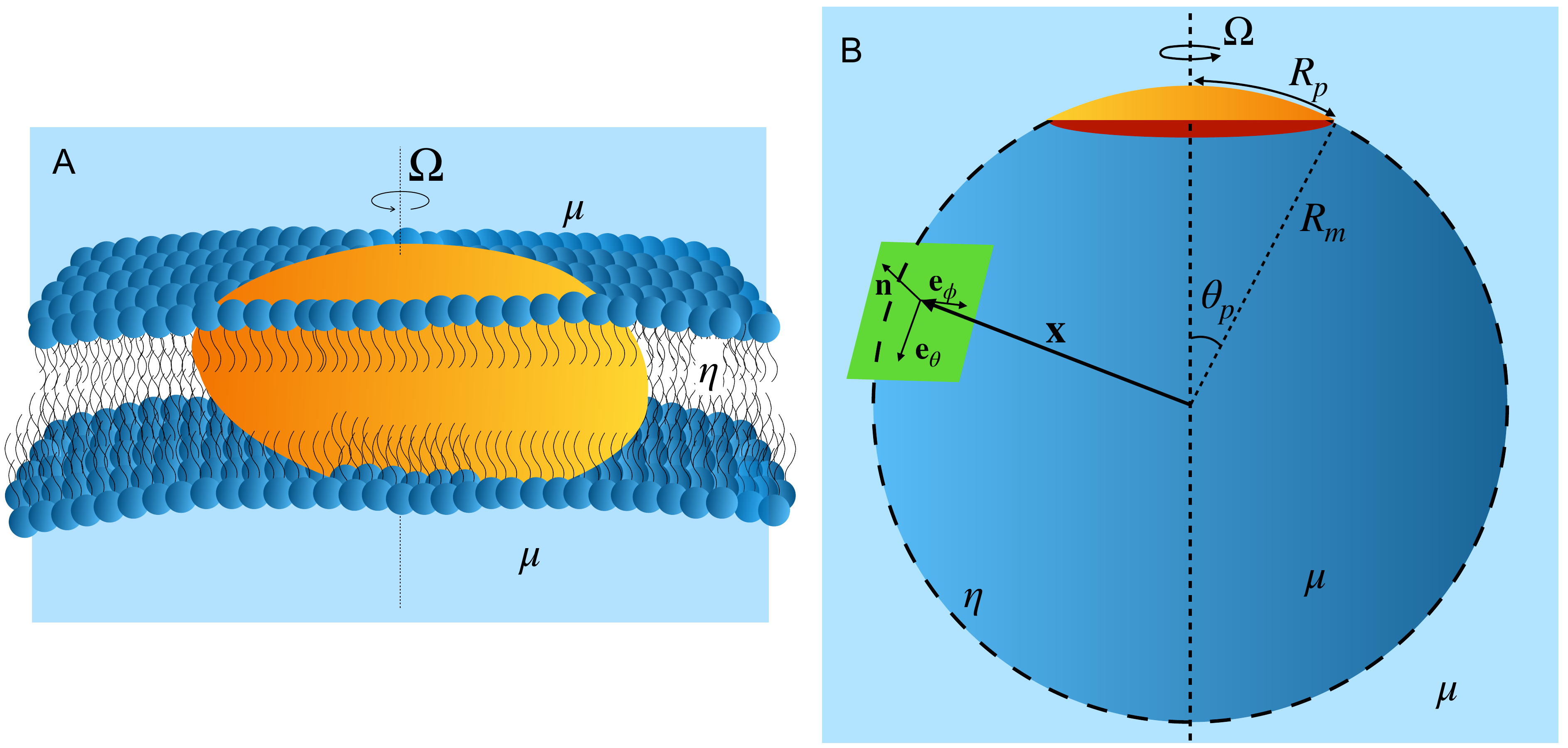}
\caption{{(A):} {Schematic of the experimental system under consideration}, the rotation of a particle  (orange structure) located inside a curved lipid bi-layer membrane (dark blue) that is embedded in a viscous solvent on both sides (light blue). {(B)}: schematic depiction of the mathematical problem. The rigid particle is modelled as a spherical cap of half-angle $\theta=\theta_p$ {and curvilinear radius $R_p$} inside a spherical membrane (vesicle) of radius $R_m$ and rotating with angular velocity $\Omega$. {The membrane is endowed with spherical polar coordinates $\theta$, $\phi$ with corresponding {orthonormal basis vectors $\mathbf e_{\theta}$, $\mathbf e_{\phi}$}}, and local unit normal $\mathbf n$.}
\label{Fig:particle Schematic}
\end{figure}

In this paper, we study the rotational mobility problem  {semi-analytically} in the case of a finite-sized rigid particle  {within} an incompressible spherical membrane.  {This is the situation illustrated schematically in Fig.~\ref{Fig:particle Schematic}A.}  Our primary motivation  concerns the rotational motion of particles embedded  {in} spherical vesicles~\cite{naji2007diffusion}, a situation  relevant to the  movement of ATP synthase~\cite{RotatingATP},  {aquaporin channels~\cite{OptimizingPermeability, WaterPermeation, AquaporinsStructure}}, {as well as} the Brownian motion of membrane-embedded particles~\cite{oppenheimer2009correlated, morris2015mobility} {, and proteins of arbitrary size~\cite{shi2022hydrodynamics, baranova2020diffusion, shi2024drag}}.
As we see below, the calculation carried out in the paper is valid  more broadly in all cases where a   rigid particle  is made to rotate inside a spherical membrane (or `vesicle'), itself embedded in a viscous solvent. {Note that our work extends the work from Ref.~\cite{henle2008effect} by allowing the particle to have a size comparable to the membrane. % This gives our theory considerable scope, by addressing the aforementioned variability in particulate size.
}

  {The structure of the paper is as follows:}  In Section~\ref{Mathematical Model}, we first outline a mathematical model for the particle-membrane-solvent system and summarise the methodology for solving the resulting equations.  In  Section~\ref{Rotational mobility of particles}, we  {focus} on the results for the  rotational mobility of the particle  and the resulting membrane flow as a function of the dimensionless parameters governing the problem (Section~\ref{sec:numerics}) and on the asymptotic values of the rotational mobility for scenarios involving a   small particle  or large particle,   comparing our findings with the results obtained in previous studies (Section~\ref{Asymptotics}). 
 We finish  {with} a summary of our findings and a discussion of potential extensions in Section~\ref{Discussion}.

\section{Mathematical Model}\label{Mathematical Model}

\subsection{Physical setup}

The motivating examples listed in Section~\ref{sec:Intro}  concern the dynamics of a rigid particle inside a spherical membrane (vesicle). The mathematical setup for our calculation is illustrated in Fig.~\ref{Fig:particle Schematic}B. The membrane forms a sphere of radius $ {R_m}$  surrounded by Newtonian solvents (both inside and outside) of  viscosity $\mu$.   The  particle is modelled as a rigid spherical cap of  {curvilinear radius $R_p$} and polar half-{angle} $\theta_p=R_p/{R_m}$, and its angular velocity  is denoted by  $\Omega$. In the case of an incompressible, {impermeable, Newtonian} membrane~\cite{arroyo2009relaxation,powers2010dynamics} of viscosity $\eta$, our  goal is to compute the total torque exerted on the particle.

{In what follows, we describe the three-dimensional space with spherical coordinates $x^i=\{r,\theta,\phi\}$. We use the standard orthonormal vectors \{$\mathbf e_r$, $\mathbf e_{\theta}$, $\mathbf e_{\phi}$\} as a local basis in the solvents, and \{$\mathbf e_{\theta}$, $\mathbf e_{\phi}$\} as a local orthonormal basis on the membrane. We assume that the membrane velocity is purely tangential, and denote the fluid velocities in the membrane and the solvents as $\mathbf v$, $\mathbf V^{\pm}$, respectively. From here onwards, a $+$ superscript denotes the exterior of the membrane, and a $-$ superscript denotes the interior.}

%These induce a local contravariant basis $\bold e_{i}=\bold r_{,i}$ and metric  {$\overline g_{ij}=\bold r_{,i}\cdot \bold r_{,j}$},  {where commas denote partial derivatives~\cite{powers2010dynamics, wald2010general}}.  The membrane corresponds to the surface $r=1$ and inherits the local basis $\bold e_{\theta}$, $\bold e_{\phi}$ and metric $g_{ab}=\text{diag}({R_m}^2,{R_m}^2\sin^2\theta)$.  {We additionally define the dimensionless outwards unit normal to the membrane $\mathbf n={\mathbf x}_{,\rho}$}. 
% {Indices of tensors in $\mathbb R^3$ are henceforth lowered and raised with the metric $\overline g_{ij}$ and its inverse $\overline g^{ij}$, while indices of tensors on the membrane are lowered and raised with the metric $g_{ab}$ and its inverse $g^{ab}$~\cite{wald2010general}}. {Assuming that the membrane velocity is purely tangential,  {we denote the} membrane and solvent velocities as $\mathbf v=v^a\mathbf e_a$ and $\mathbf V^{\pm}=\left(V^{\pm}\right)^i\mathbf e_i$ respectively  {(summation convention applies for repeated indices)},  {where a $+$ sign denotes the exterior of the membrane and a $-$ sign denotes the interior}. Notice that the $v^a$ and $(V^{\pm})^i$ have units of frequency, because $\mathbf v$ and $\mathbf V^{\pm}$ have units of velocity and {both the $\mathbf e_a$ and the $\mathbf e_i$ have units of length}. {[Removed: We henceforth non-dimensionalise the $v^a$ and $(V^{\pm})^i$ by $\Omega$.] We actually don't.}

\subsection{Mathematical model}

\subsubsection{Field equations and matching conditions}
{For an incompressible membrane with a purely tangential velocity field, mass conservation in the membrane and the solvents takes the form}~\cite{powers2010dynamics, rangamani2013interaction}
\begin{align}
{\nabla\cdot\mathbf v}&{=0,}   & {\overline{\nabla}\cdot\mathbf V=0.} \label{Incompressibility}
\end{align}
{where $\nabla$, $\overline{\nabla}$ are the gradient operators on the membrane and the solvents, respectively. In order to avoid infinite stresses, the inner and outer flows $\mathbf V^-$ and $\mathbf V^+$ must also satisfy the no-slip condition on the membrane}
\begin{align}
{\mathbf V^{\pm}}&{=\mathbf v} \quad&  {\text{(membrane)}.} \label{Membrane no-slip condition}
\end{align}
{Within the framework of continuum mechanics, forces in the membrane and the surrounding solvents (both assumed Newtonian~\cite{arroyo2009relaxation, powers2010dynamics}) are described by contravariant stress tensors $\boldsymbol{\sigma}$, $\boldsymbol{\sigma}^{\pm}$ given in an orthonormal vector basis by the constitutive relationships~\cite{powers2010dynamics, henle2008effect,arroyo2009relaxation}}
%\begin{align}
% \sigma_{ab}&=-pg_{ab}+\eta[\nabla_a \mathbf v_b+\nabla_b \mathbf v_a] & \quad &\text{{(membrane)}},\\
%{\sigma}_{ij}^{\pm}&=-P^{\pm} \overline %g_{ij}+\mu[\overline{\nabla}_i \mathbf V^{\pm}_j+\overline\nabla_j\mathbf V^{\pm}_i] & \quad&\text{{(solvents).}}
%\end{align}
\begin{align}
 {\boldsymbol{\sigma}}&{=-p\mathbf I_2^{\sharp}+\eta\left[\nabla \mathbf v+(\nabla \mathbf v)^{\text T}\right]^{\sharp}} & \quad &{\text{{(membrane)}},}\label{sigma membrane constitutive relation}\\
{\boldsymbol{\sigma}^{\pm}}&{=-P^{\pm}\mathbf I_3^{\sharp}+\mu\left[\overline{\nabla} \mathbf V^{\pm}+(\overline{\nabla} \mathbf V^{\pm})^{\text T}\right]^{\sharp}} & \quad&{\text{{(solvents).}}}
\end{align}
{Here $p, P^{\pm}$ denote the membrane and ambient pressures, {while} $\eta,\mu$ are the membrane and ambient viscosities {respectively}. A $\sharp$ denotes the index-raising sharp operator~\cite{arroyo2009relaxation}, and $\mathbf I_2$ and $\mathbf I_3$ are the $(1,1)$ identity tensors in the membrane and in the solvents.}  

{Note that in a membrane, unlike in three-dimensional space, stresses correspond to force per unit length, rather than per unit area.  As a result, $[p]=\text{Nm}^{-1}$ and $[\eta]=\text{Nsm}^{-1}$, so that {$\ell_{SD}=\eta/\mu$} has dimensions of length. The physical interpretation of $\ell_{SD}$ becomes apparent by considering an area patch of velocity $U$ and size $L$, comparable to the typical length scale of the flow. The viscous force exerted by the solvents on the patch scales as  $f_s\sim \mu UL$, while the force imparted by the membrane is of order $f_m\sim \eta U$. The drag from the solvents therefore dominates when $f_s\gg f_m\Leftrightarrow L\gg \ell_{SD}$ and the Saffman-Delbr{\"u}ck length may therefore be thought of, intuitively,  as the cross-over size between two-dimensional membrane dynamics and three-dimensional bulk dynamics~\cite{nguyen2010crossover, cicuta2007diffusion}.}

{Force balance in the solvents is expressed by the standard Stokes equation $\nabla\cdot\boldsymbol{\sigma}^{\pm}=\mathbf 0$, or}
\begin{align}
{\mu \overline{\nabla}^2\mathbf V^{\pm}=\overline{\nabla} P^{\pm}}.
\end{align}
{Similarly, we require force balance on every membrane area patch. These forces consist of membrane in-plane stresses and the traction forces $\mathbf T=(\boldsymbol{\sigma}^+-\boldsymbol{\sigma}^-)\cdot\mathbf n$ exerted by the solvents~\cite{shi2022hydrodynamics, arroyo2009relaxation}. Force balance then takes the form
\begin{align}
\text{div}_s(\boldsymbol{\sigma})+\mathbf T=\mathbf 0, \label{Force balance simplest form meain text}
\end{align}
where $\text{div}_s$ denotes the surface divergence (see Appendix~\ref{Force balance derivation}). Further decomposing $\mathbf T=\boldsymbol{\tau}+T^n\mathbf n$, with $\boldsymbol{\tau}$ tangent to the membrane, the normal and tangent components of Eq.~\eqref{Force balance simplest form meain text} may be recast into the following field equations~\cite{al2023morphodynamics, santiago2018stresses,arroyo2009relaxation}}
%\begin{align}
%\bold 0= {\text{div}_s}(\sigma^{ab}\mathbf e_a\otimes\bold e_b)+{\boldsymbol{\tau}^+}+ {\boldsymbol{\tau}^-}&=\left[\nabla_a\sigma^{ab}+(\tau^+)^b+(\tau^-)^b\right]\bold e_b+\left[K_{ab}\sigma^{ab}+(\tau^+)^n+(\tau^-)^n\right]\bold n \label{Stress Balance},
%\end{align}
\begin{align}
{\eta(\nabla^2\mathbf v+G\mathbf v)-\nabla p^{\sharp}+\boldsymbol{\tau}=\mathbf 0,}\label{Stress Balance}\\
{T^n+\boldsymbol{\sigma}:\mathbf K=0.}\label{Stress Balance 2}
\end{align}
{Here, $\mathbf K$ is the (covariant) extrinsic curvature tensor, $G=\det(\mathbf K)$ is the local Gaussian curvature, and $\nabla^2$ is the surface Laplacian. Physically, the $G\mathbf v$ term in the tangential force balance equation \eqref{Stress Balance} reflects the fact that membrane shear may occur as a result of streamlines coming together due to curvature.} 

\subsubsection{Axisymmetric solution}

{Because the setup is rotationally symmetric,  the membrane flow $\mathbf v$ and the solvent flows $\mathbf V^{\pm}$ must be everywhere parallel to $\mathbf e_{\phi}$. Indeed, the flow in the solvents cannot have any $r$ or $\theta$ component since they must change sign under reflections in a plane containing the $z$ axis (equivalent to reversing the sense of rotation of the particle). A similar symmetry argument shows that the tangential solvent stress $\boldsymbol{\tau}$ on the membrane must be purely in the $\phi$ direction, and that the normal component $T^n$ coincides with the pressure jump across the membrane, i.e.~$T^n=P^--P^+$. Finally, because the only input in this problem is the angular velocity $\mathbf\Omega$, which is a pseudo-vector, by linearity of the Stokes equations there cannot be any pressure gradients in the membrane  or the solvents. Therefore, the membrane and solvent pressures $p$ and $P^{\pm}$ must be constants.} 

{On account of these observations, only the azimuthal component of the membrane Stokes equation \eqref{Stress Balance} is non-trivial. Given the identity $\mathbf K=-R_m^{-1}\mathbf I_2^{\flat}$ in a spherical membrane with a local orthonormal basis (with $\flat$ denoting the index-lowering flat operator~\cite{arroyo2009relaxation}), the normal force balance in Eq.~\eqref{Stress Balance 2} simplifies to
\begin{align}
0&=T^n+\boldsymbol{\sigma}:\mathbf K\nonumber\\
&=P^--P^+-R_m^{-1}\boldsymbol{\sigma}:\mathbf I_2^{\flat}\nonumber\\
&=P^--P^++2R_m^{-1}p \label{Normal Stress Jump}.
\end{align}
Notice that we used the incompressibility condition $\nabla\mathbf v:\mathbf I_2^{\flat}=0$ (see Eq.~\ref{Incompressibility}) in the last step. The constant membrane pressure therefore acts like a negative tension, imposing a capillary-like pressure jump.}

{For the purpose of solving the non-trivial $\phi$ component of Eq.~\eqref{Stress Balance}, we choose to write the velocity field in a particular form, which simplifies later equations: we let $\bold v=R_m v(\theta)\sin(\theta) \bold e_{\phi}$, $\bold V^{\pm}=V^{\pm}(r,\theta) r\sin(\theta) \bold e_{\phi}$. In this formalism, the functions $v$ and $V^{\pm}$ correspond to the fluid's local  {angular velocity} about the $z$ axis, rather than the linear velocity.} Substituting {this} axisymmetric ansatz  into Eq.~\eqref{Stress Balance}, we obtain the ordinary differential equation
\begin{align}
{\eta(v''\sin \theta+3v'\cos\theta)+ {R_m}\tau^{\phi}}&{=0} &\theta_p<\theta\leq\pi{\text{ (membrane)}}\label{NS phi direction},\\
v& \equiv  {\Omega} & 0\leq \theta\leq \theta_p{\text{ (particle)}}.\label{NS Boundary Condition}
\end{align}
Note that this equation may also be obtained by considering the standard Stokes equation in a thin spherical annulus, assuming that the thickness-integrated body force balances with external shear  {(see Appendix~\ref{Membrane Flow Equation as Limit of the $3$D Stokes Equations})}.

\subsection{Legendre polynomial expansion}
Following  the  classical  {squirmer} solution~\cite{SquirmingSolution} we decompose  the three flows  {in a basis} of Legendre polynomials. Writing  $x\coloneqq\cos\theta$,  {we use $P_n(x)$ to denote to the $n$-th Legendre polynomial and $P_n'(x)$   its derivative with respect to $x$}. We look for {three solutions of the forms}~\cite{SquirmingSolution}
\begin{align}
 V^+&={\sum_{n=1}^{\infty}c_n \left(\frac{R_m}{r}\right)^{n+2}P'_n(x)} &  {P^+}& {\equiv  {q_0}} \label{Outer Solution} & r&>{R_m},\\
V^-&={\sum_{n=1}^{\infty}c_n \left(\frac{r}{R_m}\right)^{n-1}P'_n(x)} &  {P^-}& {\equiv  {q_0}-\frac{2 p_0}{R_m}} & r&<{R_m}\label{Inner Solution},\\
v&=\sum_{n=1}^{\infty}c_nP_n'(x) &  {p}& {\equiv p_0} & r&={R_m}\label{Expansion to Test Against},
\end{align}
where we accounted for the normal stress jump~\eqref{Normal Stress Jump},  {the no-slip condition at $r=R_m$}~\eqref{Membrane no-slip condition}, and  {enforced} regularity at both $r=0$  {and} $r\to\infty$.    {Note that, in principle, we should have three independent sets of expansion coefficients $a_n$, $b_n$ and $c_n$ for each of $V^+$, $V^-$, $v$ in Eqs.~(\ref{Outer Solution}), (\ref{Inner Solution}), and (\ref{Expansion to Test Against}). However, by the no-slip condition in Eq.~(\ref{Membrane no-slip condition}), all three  {expansions} must coincide when {$r=R_m$}. Because the $P'_n$ are orthogonal with respect to the inner product,
\begin{align}
\langle P'_n,P'_m\rangle\coloneqq \int_{-1}^1 P_n'(x)P_m'(x)(1-x^2)\mathrm dx,
\end{align}
the coefficients of all three expansions must  {be identical}.}

\begin{center}
\setlength{\tabcolsep}{20pt} 
\renewcommand{\arraystretch}{1}
\begin{table}[t]
\begin{tabular}{ c c c }
 Coefficient & Velocity & Interpretation \\\hline 
 $c_1$ & $r^{-2}\sin\theta$ & rotlet\\  
 $c_2$ & $r^{-3}\sin 2\theta$ & rotlet dipole\\
 $c_3$ & $r^{-4}(5\cos^2\theta-1)\sin\theta$ & rotlet quadrupole\\
 \vdots & \vdots & \vdots
\end{tabular}
\caption{{Velocity field corresponding to each expansion coefficient in Eqs.~\eqref{Outer Solution}. The coefficient $c_{n+1}$ corresponds to an rotlet $2^{n}$-pole, obtained by differentiating the rotlet flow $n$ times with respect to $z$.}}\label{Rotlet multipole table}
\end{table}
\end{center}

 {The set of coefficients $c_n$ represent the  strengths of {rotlet moments} of progressively higher order (rotlet, rotlet dipole etc.). To see this, let us consider the first few terms of the expansion \eqref{Outer Solution} and list in Table \ref{Rotlet multipole table} (up to multiplicative constants) the corresponding {contributions to the external azimuthal flow component $\mathbf V^+\cdot\mathbf e_{\phi}$.} 
 {Each tabulated} singularity is the gradient of the previous one along the $z$ axis. {A pictorial representation of the flows associated with the first three moments is provided in Fig.~\ref{Fig:Rotlet Moments}.} Intuitively, the  {far-field} flow {outside the membrane} is expected to be composed of a contribution from solid-body motion (rotlet), a contribution from  {the differing rotations} of the particle and the fluid membrane due to external drag (rotlet dipole) and higher-order singularities capturing how the drag is distributed on the surface.
 
  \begin{figure}[t]
   \centering
\includegraphics[width=1\textwidth]{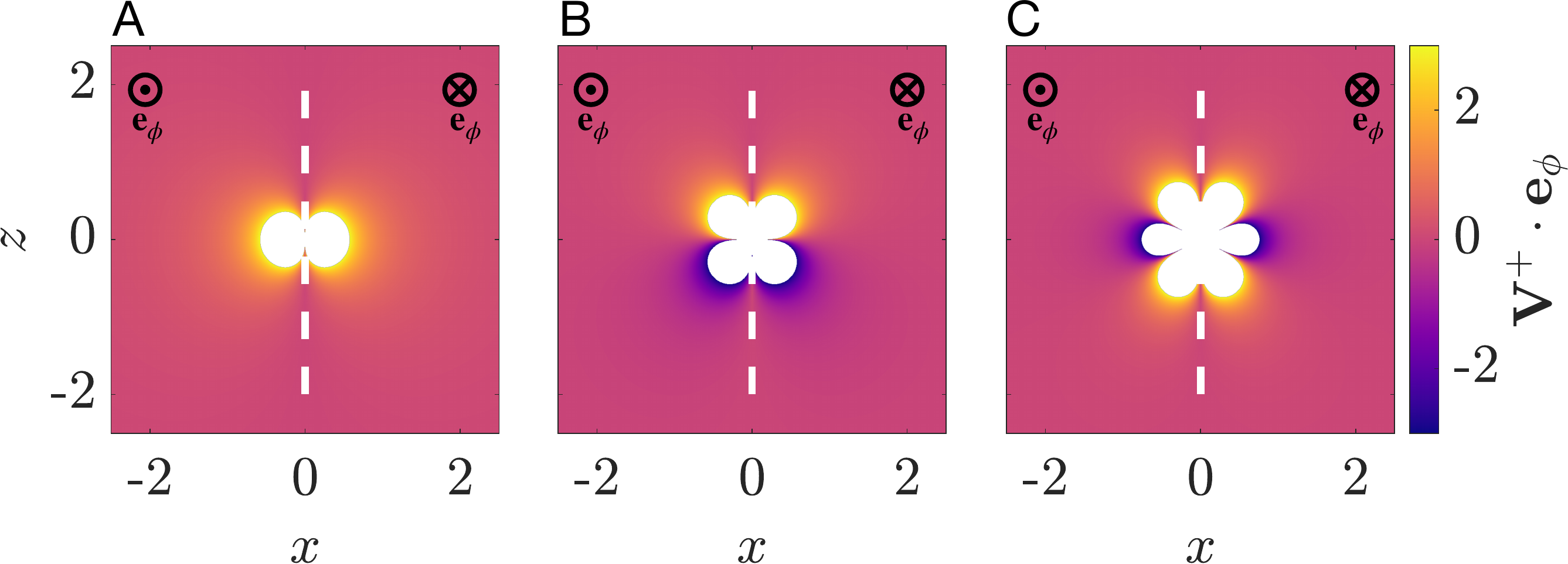}
\caption{{Flows associated with (A) a rotlet, (B) a rotlet dipole, (C) a rotlet quadrupole located at the origin and oriented long the $z$ axis (dashed white line). In each case, the flow rotates around the $z$ axis, so a single slice is plotted corresponding the the $xz$ plane. The boundaries of the white patches mark the region where the velocity magnitude (infinite at the origin) first exceeds a set threshold.}}
\label{Fig:Rotlet Moments}
\end{figure}
 
 We will show in what follows that the total torque on the particle  {is directly captured by} the rotlet strength $c_1$, as would be expected from physical intuition.}
To fully characterise the flows, we  {need} to determine the  coefficients $c_n$.  {With $x=\cos\theta$ and exploiting orthogonality of the Legendre polynomials, the force-balance and no-slip equations~\eqref{NS phi direction},~\eqref{NS Boundary Condition},~\eqref{Expansion to Test Against} can be expressed as (see Appendix~\ref{Legendre Polynomial Expansion})}
\begin{align}
&\frac{(n+1)(n+2)}{(2n+1)(2n+3)}c_{n+1}-\frac{n(n-1)}{(2n-1)(2n+1)}c_{n-1}=\frac{1}{2}\int_{-1}^1P_n(x)v(x)(1-x^2)\mathrm dx & &n\geq 0,  \label{c_n Recursion}\\
&(1-x^2)v''(x)-4xv'(x)=\frac{\varepsilon}{2\theta_p}\sum_{n=1}^{\infty}(2n+1)c_nP_n'(x) & x&<x_p\label{Navier Stokes Expanded Forcing1},\\ 
 &v(x)\equiv  {\Omega} & x&\geq x_p\label{Navier Stokes Expanded Forcing2}.
\end{align}
Here, {$x_p=\cos\theta_p$}. {Furthermore}, $\varepsilon=2\mu R_p/\eta$ is the ratio between the width of the particle,   {$R_p=R_m\theta_p$}, and the Saffman-Delbr{\"u}ck length, $\ell_{\text{SD}}=\eta/\mu$~\cite{CylinderInMembrane}. The coefficient appearing on the right-hand side of  Eq.~\eqref{Navier Stokes Expanded Forcing1} is thus given by $\varepsilon/2\theta_p=R_m/\ell_{\text{SD}}$,  {the} ratio between the radius of curvature of the membrane and the Saffman-Delbr{\"u}ck length.

{Finally, note that the membrane velocity $\mathbf v$ has magnitude $\lVert\mathbf v\rVert=R_m|v|(1-x^2)^{1/2}$.} 
We may thus allow singularities in $v(x)$  {at $x=\pm 1$} provided that the velocity is continuous, i.e.
\begin{align}
v(x)(1-x^2)^{1/2}\rm{\quad continuous} \label{Continuity Condition}.
\end{align}

 {Therefore,} the flow is  {fully} determined by Eqs.~(\ref{c_n Recursion})-(\ref{Continuity Condition}). These equations also show that the dimensionless flow {$v/\Omega$} is only a function of two dimensionless parameters, $\varepsilon$ and $\theta_p$. 
 
\subsection{Rotational mobility}
The rotational mobility is obtained by computing the  total torque {$G_p\mathbf e_z$} exerted on the particle {to maintain the rotation}. {This torque must balance from the drag exerted by the solvents on the top and the bottom of the particle, as well as the in-plane drag on the particle's edge due to the membrane. As shown in Appendix \ref{Force balance derivation}, however, the total torque on the particle's edge is precisely equal to the total torque on the membrane with $\theta_p\leq\theta\leq \pi$. We deduce that $G_p$ must balance to the total viscous torque on the particle-membrane assembly caused by the solvents.} In other words
\begin{align}
{G_p}&{=-2\pi R_m^3\int_{-1}^1 {\tau^{\phi}}(1-x^2)^{1/2}\mathrm dx}\nonumber\\
&=2\pi \mu R_m^3\sum_{n=1}^{\infty}(2n+1)c_n\int_{-1}^1P_n'(x)(1-x^2)\mathrm dx.
\end{align}
All terms in  {the integral above} vanish except the first one, leading to
\begin{equation}
G_p=8\pi\mu c_1R_m^3, \label{Torque c1 relation}
\end{equation}
 {and} thus all information about the torque is embedded in the rotlet coefficient $c_1$, as {expected}. In the rest of the paper, we will write the drag using a standard normalisation  {in terms of} the particle radius $R_m\theta_p$ as 
 $G_p\equiv 8\pi \mu  {\Omega} R_m^3\theta_p^3\Lambda_R$, thereby introducing the dimensionless drag coefficient $\Lambda_R(\varepsilon,\theta_p) =  {\Omega^{-1}}\theta_p^{-3}c_1  $.

  \begin{figure}[t]
\includegraphics[width=1\textwidth]{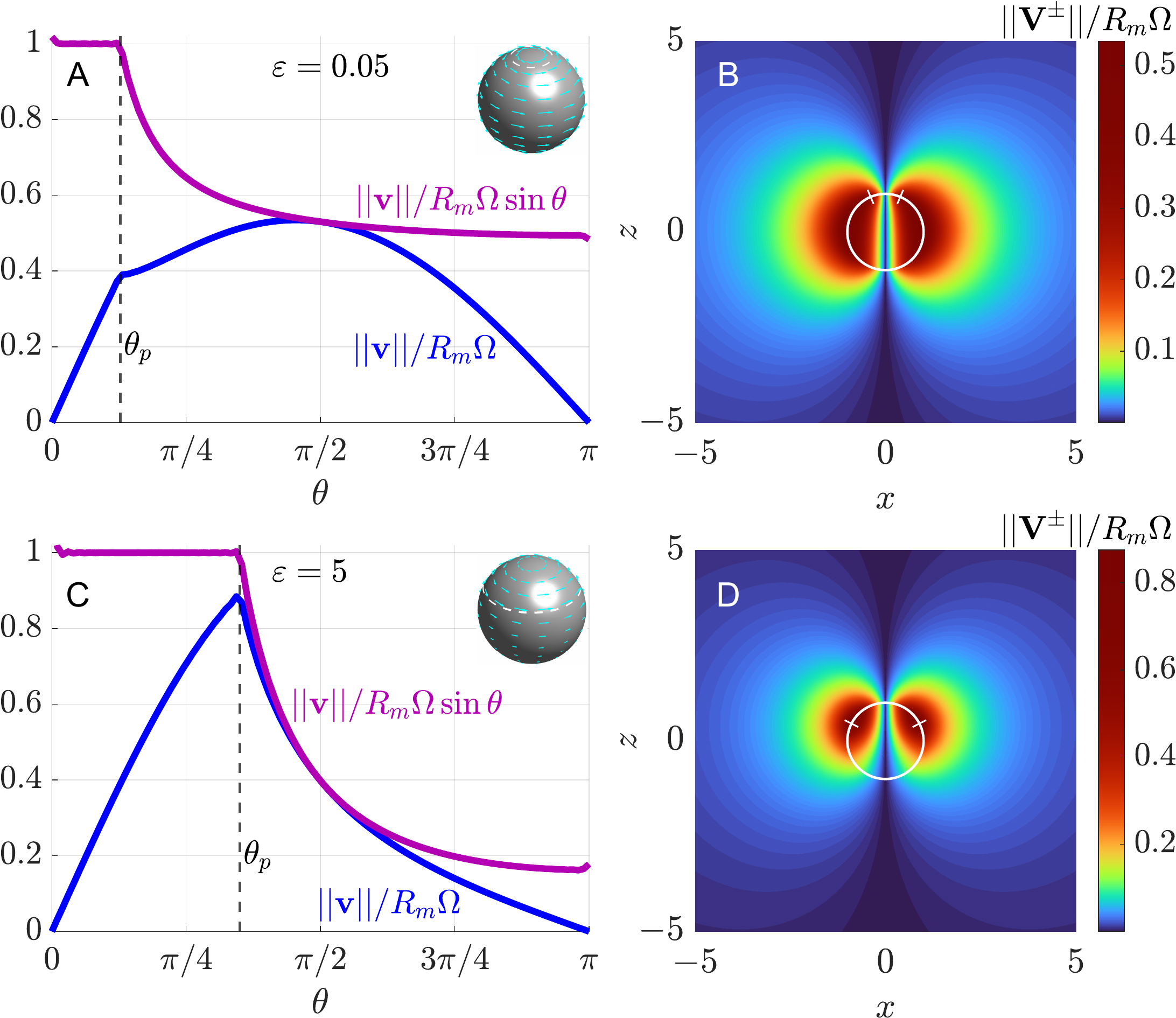}
\caption{{Flows within the membrane and the solvents for different inclusion sizes and relative Saffman lengths $\varepsilon=2R_p/\ell_{SD}$. (A) Normalised flow in the membrane for $\varepsilon=0.05$, $\theta_p=0.4$ (blue line) and its relative solid-body motion component (purple line). Inset: $3$D plot of the membrane velocity field. (B) Illustration of solvent flow (iso-magnitude) for the same values of $\varepsilon$, $\theta_p$ as (A); since the setup has rotational symmetry, we only  {plot} the magnitude of the azimuthal flow component of the ambient flows in the plane $y=0$. The white circles represent the  {membrane}, with  {radial} ticks marking the particle edge. (C), (D): same as (A), (B) with parameters $\varepsilon=5$, $\theta_p=1.1$}}
\label{Fig:Ambient Flows particle}
\end{figure}

\section{Rotational mobilities and membrane flow}
\label{Rotational mobility of particles}
The dimensionless drag coefficient $\Lambda_R$ is a function of two dimensionless parameters: $\varepsilon=2\mu R_p/\eta$ (ratio of
particle size and the Saffman-Delbr{\"u}ck length~\cite{henle2008effect}) and the dimensionless particle  {half-angle $\theta_p$}.

\subsection{Numerical results}
\label{sec:numerics}
 We solve the problem  by truncating all sums at $n=k$ for some finite $k$ and then computing the solutions to Eqs.~(\ref{c_n Recursion}), (\ref{Navier Stokes Expanded Forcing1}), (\ref{Navier Stokes Expanded Forcing2}),  (\ref{Continuity Condition})  {semi-analytically}.  {We first solve Eq.~(\ref{Navier Stokes Expanded Forcing1}) for $v(x)$ in terms of the expansion coefficients $c_n$, and obtain a set of coupled, linear algebraic equations for $c_1$,..., $c_k$ from Eqs.~(\ref{c_n Recursion}), (\ref{Navier Stokes Expanded Forcing2}). These can then be solved to find $c_1$, and hence $\Lambda_R$ (Eq.~\ref{Torque c1 relation})}.   {Numerically, we see that the number of modes required to attain a given accuracy diverges as $\theta_p\to 0$. We henceforth use $k=100$ for $\theta_p\geq 0.4$, $k=200$ for $0.2\leq\theta_p<0.4$ and $k=300$ for $\theta_p<0.2$. These values of $k$ ensure that, in the limit $\eta\to 0$, we recover the mobility of a rotating spherical cap with a relative error no larger than $2\%$ (see \S~\ref{Low curvature limit})}. With the coefficients known, we can evaluate the flow in the membrane and in both solvents, and deduce the rotational mobility of the particle.

We first  illustrate in Fig.~\ref{Fig:Ambient Flows particle} the  {dimensionless} magnitude of the membrane velocity,  {$\lVert\mathbf v\rVert/R_m\Omega$} {and angular velocity $\lVert\mathbf v\rVert/R_m\Omega\sin\theta$, as well as} the  {dimensionless} magnitude of the ambient  {flows $\lVert\mathbf V^{\pm}\rVert/R_m\Omega$} for two representative choices of the parameters $\varepsilon$ and $\theta_p$: 
$\varepsilon=0.05$, $\theta_p=0.4$ (A-B) and $\varepsilon=5$, $\theta_p=1.1$ (C-D).  {These values were inspired by experiments with phase-separated giant unilamellar vesicles~\cite{cicuta2007diffusion}. {Taking the representative values $R_p\sim 1-10\ \mu$m, $\ell_{SD}\sim 10-10^3\ \mu$m~\cite{cicuta2007diffusion} provides the estimate $10^{-3}\leq \varepsilon\leq 10$}. In all cases:  {(i) the particle does rotate like a rigid body, as prescribed {(i.e.~with constant angular velocity)};} (ii) the rotational velocity also vanishes at both the north and south pole, as imposed by regularity; and (iii) the external velocity field decays like a rotlet, i.e.~{as} $\mathcal O(r^{-2})$. Outside the cap, the velocity profile has an internal maximum when $\varepsilon/2\theta_p {=R_m/\ell_{\text{SD}}} \ll 1$ (i.e.~in the limit of a very viscous membrane), by analogy with a rotating solid sphere~\cite{samanta2021vortex}  (Fig.~\ref{Fig:Ambient Flows particle}, A-B). Mathematically, this limit corresponds e.g.~to sending $\eta\to\infty$ for fixed $\theta_p$. Quantitatively, note that sending $\varepsilon/2\theta_p\to 0$ turns Eq.~(\ref{Navier Stokes Expanded Forcing1}) into
\begin{align}
(1-x^2)v''-4xv'=0  
\end{align}
for $-1\leq x\leq x_p$, whose only regular solution is $v\equiv  {\Omega}$. In  {this} limit, {the flow in Fig.~\ref{Fig:Ambient Flows particle}A therefore approaches rigid-body motion, corresponding to a perfect sine wave}. {This effect is particularly evident in the angular velocity plot, which remains fairly close to $1$ throughout}. Conversely, when  {$\varepsilon/2\theta_p\gtrsim\mathcal O(1)$}, the membrane flow is seen to decay monotonically  (Fig.~\ref{Fig:Ambient Flows particle}, C-D), {and the angular velocity is substantially smaller than for rigid-body motion.}

Next, we study the dependence of the {torque exerted on the particle} on the two relevant dimensionless parameters, $\varepsilon$ and $\theta_p$. The results are displayed in  Fig.~\ref{Fig:Drag Coefficient Many Values}{A}, where we plot the torque $G_p$ applied on the particle to maintain the rotation as a function of the particle half-angle $\theta_p$,  non-dimensionalised by the torque $G_m$ exerted on a rigid sphere  {of radius $R_m$} rotating with the same angular velocity.  {The chosen parameter values, $\theta_p\geq 0.1$ and $0.1\leq \varepsilon\leq 100$, correspond to a membrane viscosity of $10^{-10}~\text{Nsm}^{-1}\lesssim \eta\lesssim 10^{-6}~\text{Nsm}^{-1}$} for a $50\ \mu$m vesicle, a range covering typical values for giant unilamellar vesicles~\cite{cicuta2007diffusion}.

   \begin{figure}[t]
   \centering
\includegraphics[width=1\textwidth]{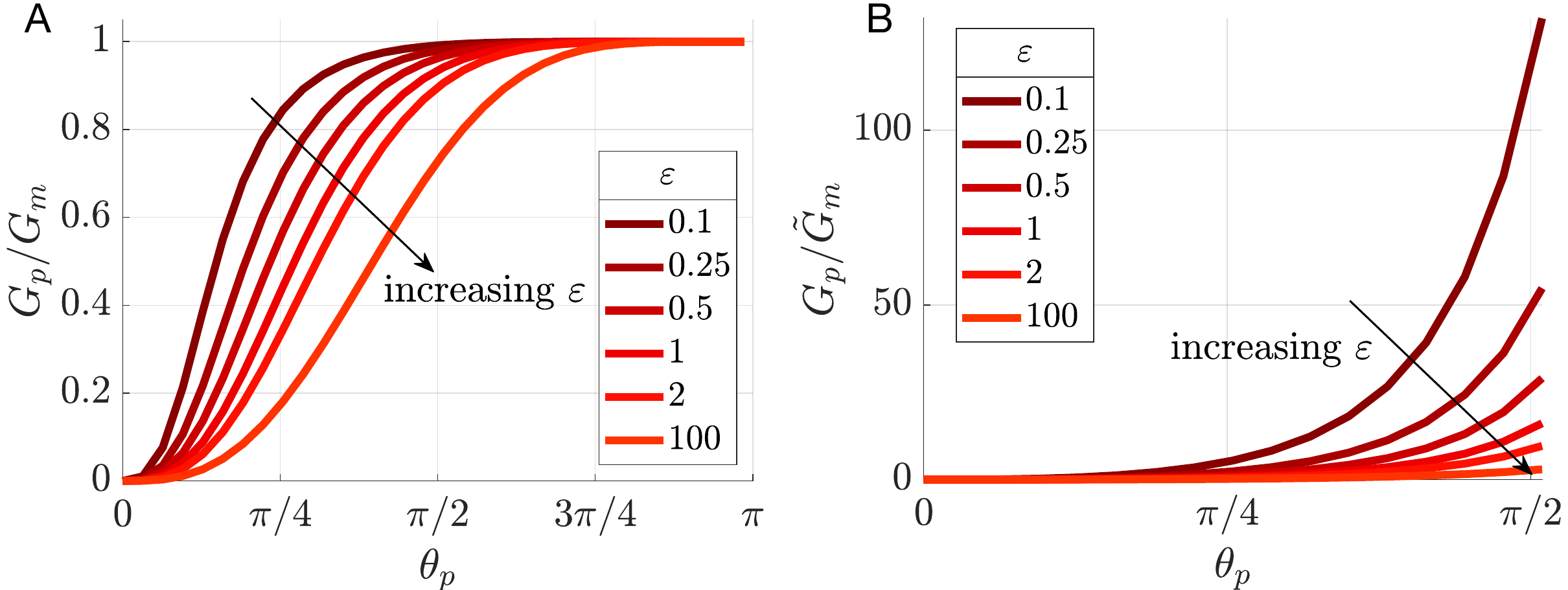}
\caption{{Dependence of the {exerted torque} on the particle size and the membrane viscosity:} (A) torque $G_p$ exerted on the  {particle} to maintain rotation (non-dimensionalised by the torque  {$G_m$} on a rigid sphere of radius $R_m$ rotating with the same angular velocity  {in {a solvent} with viscosity $\mu$}) as a function of  the half-{angle}  {$\theta_p$} of the  {particle}, for  {a range of} values of the modified viscosity ratio {$\varepsilon=2\mu R_p/\eta=2R_p/\ell_{SD}$. For a fixed particle size, the torque increases for larger $\ell_{SD}$ (i.e.~{for decreasing} $\varepsilon$). {(B) Torque $G_p$ exerted on the particle to maintain rotation non-dimensionalised by the torque {$\tilde G_m$ exerted } on a rigid sphere (radius $R_m$, ambient viscosity $\mu$) rotating at the modified rate $\Omega-c_1$ (i.e.~the difference between $\Omega$ and the rigid motion of the membrane).}}}
\label{Fig:Drag Coefficient Many Values}
\end{figure}

As expected, the normalised torque  {$G_p/G_m\to 1$} as  {$\theta_p\to \pi$} since, in this limit, the particle covers the entirety of the vesicle, making the membrane completely rigid. Conversely, the torque vanishes for a {small particle ($\theta_p\to 0$)}. {For a given particle size and solvent viscosities, the applied torque  increases with the Saffman-Delbr{\"u}ck  length $\ell_{SD}$, as  {a larger $\ell_{SD}$ corresponds to a more viscous, and hence more rigid, membrane}.} This effect is more obvious for larger values of  {$\theta_p$} as the membrane's angular velocity, which is nearly constant due to the small azimuthal shear, has less room to vary.  {Finally, in Fig.~\ref{Fig:Drag Coefficient Many Values}B we explore a different definition of particle mobility~\cite{henle2008effect}, defined in the frame co-rotating with the rigid-body motion of the membrane. In particular, we plot the torque on the inclusion non-dimensionalised by the torque $\tilde G_m$ on a rigid sphere (radius $R_m$) rotating with angular velocity $\Omega-c_1$ in the same solvents, where $c_1$ corresponds to the membrane's solid-body motion.
The ratio $G_p/\tilde G_m$ is larger when the membrane is more viscous (smaller $\varepsilon$), as this impedes differential rotation between the  membrane and the particle. The relative torque in Fig.~\ref{Fig:Drag Coefficient Many Values}B also blows up for large inclusions ($\theta\to\pi$) as shearing the membrane becomes increasingly hard. A co-rotating drag coefficient may then be defined as $\tilde\Lambda_R=(\Omega-c_1)^{-1}\theta_p^{-3} c_1$. In terms of the free mobility, $\tilde \Lambda_R=(1-c_1/\Omega)^{-1}\Lambda_R$, implying that the two coefficients differ when the solid-body rotation of the membrane is significant (i.e.~for sufficiently large $\ell_{SD}$ or large inclusions). }

\subsection{{Asymptotic Results}}\label{Asymptotics}
 Our setup contains three inherent length scales, namely the curvilinear particle radius $R_p$, the membrane radius $R_m$, and the Saffman-Delbr{\"u}ck length $\ell_{SD}$, with corresponding dimensionless ratios $R_p/R_m=\theta_p$ and $\ell_{SD}/R_m=2\theta_p/\varepsilon$. To understand the  interactions between the various length scales, we now analyse two asymptotic limits, with results summarised in Fig.~\ref{Fig: Asymptotics}. First,  {in \S~\ref{Small particle Limit}}, we consider the limit  of a small particle  {($R_p/R_m\ll 1$ or $\theta_p\ll 1$)} of varying Saffman-Delbr{\"u}ck length,  {or equivalently of varying $\varepsilon$}. In this case, we observe a transition from planar mobility~\cite{CylinderInMembrane} to spherical mobility as the membrane becomes more and more {viscous} ($\ell_{SD}\to\infty$). We then address the limit of a particle of size comparable to the membrane, i.e.~{$R_p\sim R_m$} (\S~\ref{Low curvature limit}); we demonstrate that, in the limit of small Saffman-Delbr{\"u}ck length  {($\ell_{SD}\ll R_p$ or $\varepsilon\gg 1$)}, the hydrodynamic effects of the membrane are negligible and the particle experiences the same drag as a rotating spherical cap in an unbounded fluid~\cite{SphericalCapTorque}. {As the Saffman length increases, the torque instead becomes the same as on a rotating rigid sphere (see Fig.~\ref{Fig:Drag Coefficient Many Values}).}

 \begin{figure}[H]
\centering
\includegraphics[width=0.5\textwidth]{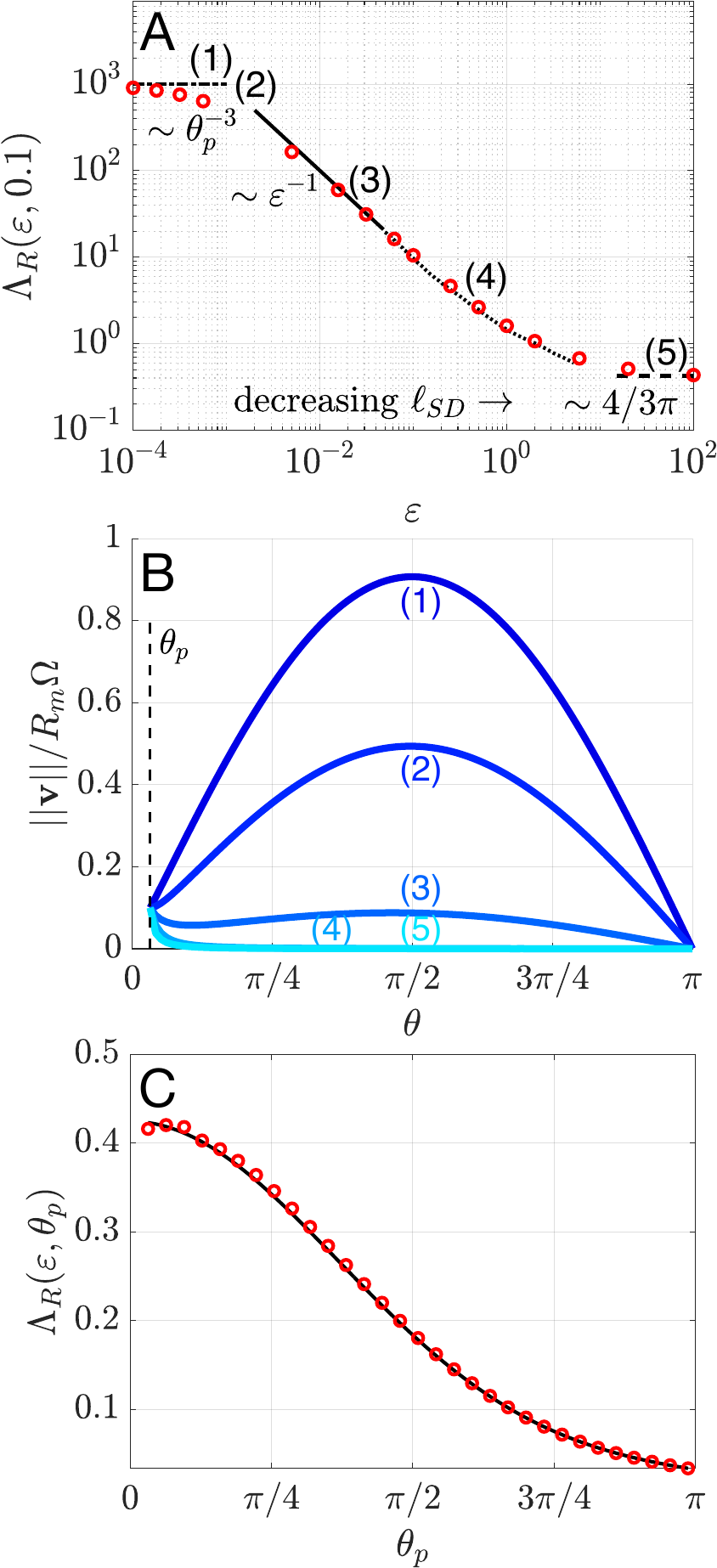}
\caption{Asymptotic behaviour of rotational drag coefficient, $\Lambda_R(\varepsilon,\theta_p)$, in the limits of small and large particles. A: Small particle limit, $R_p\ll R_m$ (or $\theta_p\ll 1$) for {many} values of $\varepsilon=2R_p/\ell_{SD}$ (numerics  run with $\theta_p=0.1$, $k=300$), with details in    \S~\ref{Small particle Limit}. The numerical values of $\Lambda_R$ (red circles) are compared with {(1)} the solid-body motion asymptotic $\theta_p^{-3}$, {(3)} the asymptotics $\Lambda_R\sim \varepsilon^{-1}$ (solid black line), {(4)} the analytical prediction from Fig.~$3$ in Ref.~\cite{CylinderInMembrane} (dotted black line), and {(5)} the asymptotics $\Lambda_R\sim 4/3\pi$ for a rotating disc~\cite{FluidsBook} (dashed black line). \ {Region {(2)} marks the transition between the three-dimensional mobility in {(1)} and the nearly-planar mobility in {(3)}.}  {B: Plot of the dimensionless membrane velocity ($\theta\geq \theta_p$) for a small particle ($\theta_p=0.1$) and representative values of $\varepsilon$ for each region: {(1)} $\varepsilon=10^{-4}$, {(2)} $\varepsilon=10^{-3}$, {(3)} $\varepsilon=10^{-2}$, {(4)} $\varepsilon=2$, {(5)} $\varepsilon=20$.} C: Small Saffman-Delbr{\"u}ck length limit, $\ell_{SD}\ll R_p\sim R_m$ (or $\theta_p/\varepsilon\ll \theta_p, 1$), with details in  \S~\ref{Low curvature limit}. Numerical results ($0.1\leq \theta_p\leq 3.1$ and $\varepsilon=200$, red circles) compared with the analytical solution for a spherical cap ({solid black line})~\cite{SphericalCapTorque}. %{[Figure was edited to improve readability]}
}
\label{Fig: Asymptotics}
\end{figure}

\subsubsection{{Small particle limit: $R_p/R_m\ll 1$}}\label{Small particle Limit}

{Firstly, we consider a particle that is much smaller than the membrane and compare the numerically calculated mobility with the planar value~\cite{CylinderInMembrane}.   {Experimentally, this limit is appropriate for smaller inclusions, such as membrane-bound proteins and microspheres~\cite{LipidBilayer}. It is common in the literature to infer the membrane's viscosity by measuring the diffusion coefficient of such inclusions and relating it to the viscosity as in Saffman's planar theory~\cite{SaffmanFlow, SaffmanDetailedCalculations}. We will show below that, in the rotational case, such approximation is appropriate as long as $\eta$ is not too large.}
As the Saffman-Delbr{\"u}ck length is varied from small values (highly viscous solvents) to large values (highly viscous membrane), five distinct asymptotic regions emerge (Fig.~\ref{Fig: Asymptotics}A), which we set out to explain below.}

{\paragraph{Rigid rotation {(1)}.}

 When $\ell_{SD}$ is large (Fig.~\ref{Fig: Asymptotics}A, region {(1)}), the membrane rotates rigidly to minimise dissipation (Fig.~\ref{Fig:Drag Coefficient Many Values})~\cite{henle2008effect, samanta2021vortex}. The torque on the cap is therefore approximately $\sim 8\pi\mu \Omega R_m^3$, and therefore $\Lambda_R\sim \theta_p^{-3}$. This regime occurs when the torque associated with rigid rotation of the whole vesicle is much less than the torque associated with non-zero membrane shear. {Importantly, this regime is not captured by the planar Saffman theory, which assumes the membrane to be at rest far from the inclusion.}

\paragraph{Transition {(2)} and two-dimensional limit {(3)}.}

{As the Saffman-Delbr{\"u}ck length is reduced, the membrane begins to experience   shear from the particle. For $\ell_{SD}\gg R_p$, the drag from the solvents is negligible on the scale of the particle and the local flow is purely two-dimensional~\cite{nguyen2010crossover}. The particle mobility is in this case $\Lambda_R\sim \varepsilon^{-1}$, the same as a disk in an infinite two-dimensional fluid without any solvents, a problem that is not subject to Stokes paradox (Fig.~\ref{Fig: Asymptotics}A, region {(3)}) ~\cite{FluidsBook, SaffmanDetailedCalculations, CylinderInMembrane}. The transition region {(2)} between the previous two regimes therefore occurs when $\varepsilon^{-1}\sim \theta_p^{-3}$, signifying that regime {(1)} corresponds to $\varepsilon \ll \theta_p^3$, or equivalently $\ell_{SD}\gg R_m^3/R_p^2$. Region {(3)} instead corresponds to $\theta_p^3\ll \varepsilon\ll 1$, or $R_p\ll\ell_{SD}\ll R_m^3/R_p^2$.  {Region {(2)} is physically significant as the inclusion ceases to feel the effect of the membrane's geometry (dominant in {(1)}) and the flow becomes local, rendering the mobility nearly planar (region {(3)}).}}

{The  {asymptotic} behaviour in region {(3)}, corresponding to $\theta_p^3\ll \varepsilon\ll 1$ or $R_p\ll \ell_{SD}\ll R_m^3/R_p^2$,  {can be recovered mathematically} by noting that the local flow varies on the length scale of the particle~\cite{samanta2021vortex}, since the solvents are negligible within a distance $\mathcal O(\ell_{SD})$ of the particle.  {Therefore,} the torque is caused predominantly by the in-plane membrane shear, implying that
%\begin{align}
%\Lambda_R&\sim -\frac{1}{8\pi \mu R_m^3\theta_p^3}\int_{\phi=0}^{2\pi} \sigma^{\phi}_{\theta}(\bold e_r\times \bold e_{\phi})\cdot\bold e_z \sin\theta\mathrm d\phi=-\frac{1}{8\pi \mu R_m\theta_p^3}\int_{\phi=0}^{2\pi}\sigma^{\theta}_{\phi}\sin\theta \mathrm d\phi,
%\end{align}
%and, thus,
%\begin{equation}\label{eq:High curvature limit}
%\Lambda_R \sim\frac{v'(x_p)(1-x_p^2)^2}{2\varepsilon \theta_p^2 {\Omega}}.
%\end{equation}
\begin{align}
{\Lambda_R}&{\sim \frac{1}{8\pi \mu \Omega R_m^3\theta_p^3}\int_{\phi=0}^{2\pi} R_m^2\mathbf e_r\times\boldsymbol{\sigma}\cdot\mathbf e_{\theta}\sin \theta_p\mathrm d\phi= \frac{v'(x_p)(1-x_p^2)^2}{2\varepsilon \theta_p^2 {\Omega}}.}
\end{align}
To proceed, we need to determine $v$ around the particle. Since the flow varies on an angle $\theta_p$, by locally writing $x=1-\theta_p^2X$ with $X=\mathcal O(1)$, from Eq.~(\ref{Navier Stokes Expanded Forcing1})
\begin{equation}
\begin{aligned}
&(1-x^2)v''(x)-4xv'(x)=\mathcal O(R_mv/\ell_{SD})\\
 {\Rightarrow}\ & {Xv''(X)+2v'(X)=\mathcal O(R_p^2v/R_m\ell_{SD})\sim 0} \label{Asymptotics intermediate step}\\
\Rightarrow\ &v=\frac{A}{X}+B,
\end{aligned}
\end{equation}
where in Eq.~{(\ref{Asymptotics intermediate step})} we exploited the fact that $R_p^2v/\ell_{SD}R_m\ll v$.}

{Notice that the torque associated with the solid-body motion $B$ is $\sim \mu BR_m^3$, while the torque on the particle is $G_p\sim \eta R_p^2 v$. Equating these, we obtain $B\sim \ell_{SD}R_p^2 v/R_m^3$, which is much less than $v$ in region {(3)}. Setting $B=0$ at leading order, the boundary conditions are $v= {\Omega}$ on $x=1-\theta_p^2/2+\mathcal O(\theta_p^4)$, i.e.~$v(X=1/2)= {\Omega}$. This yields $A= {\Omega}/2$, giving the inner solution
\begin{align}
v=\frac{ {\Omega}}{2X}=\frac{ {\Omega}(1-x_p^2)}{2(1-x)}.    
\end{align}
This is indeed analogous to the flow around a spinning cylinder in a flat, solvent-free membrane  {($v\sim d^{-2}$ with $d=\sin\theta/\sin\theta_p$)}. We deduce that, as $x_p\to 1$, 
\begin{align}
v'(x_p)(1-x_p^2)\approx \frac{1}{2} {\Omega}(1+x_p)^2\approx 2 {\Omega},
\end{align}
and therefore the dimensionless friction coefficient is
\begin{align}
 \Rightarrow \Lambda_R\sim \varepsilon^{-1} \label{Lambda R Asymptotic 1},
\end{align}
as observed in Fig.~\ref{Fig: Asymptotics}A.}
 
\paragraph{Planar regime {(4)}.}

{As the membrane viscosity is reduced further, eventually $\varepsilon\sim 1$ or $\ell_{SD}\sim R_p$ (Fig.~\ref{Fig: Asymptotics}A, region {(4)}). In this, regime the mobility is asymptotically the same as in the planar regime with {contributions to the drag from both the membrane and the solvents}~\cite{CylinderInMembrane}, as the small particle is oblivious to the membrane's geometry.}

\begin{center}
\setlength{\tabcolsep}{8pt} 
\renewcommand{\arraystretch}{1}
\begin{table}[t]
\begin{tabular}{ c |c |c | c}
Asymptotic limit & Drag Coefficient & Refs & Interpretation\\\hline
$\ell_{SD}\gg R_m^3/R_p^2$  & $\Lambda_R\sim (R_m/R_p)^3$ & \cite{kim2013microhydrodynamics, happel2012low} & 
 Solid-body membrane rotation\\
$\ell_{SD}\sim R_m^3/R_p^2$ & $\Lambda_R\sim (R_m/R_p)^3$ & \cite{kim2013microhydrodynamics, happel2012low} & Near-rigid membrane rotation with small shear\\
$R_p\ll \ell_{SD}\ll R_m^3/R_p^2$ & $\Lambda_R\sim \ell_{SD}/2R_p$ & \cite{SaffmanFlow} & Cylinder in $2$D fluid (no solvents)\\
$\ell_{SD}\sim R_p$ & $\Lambda_R=\mathcal O(1)$ & \cite{CylinderInMembrane} & Cylinder in $2$D fluid with solvents\\
$\ell_{SD}\ll R_p$ & $\Lambda_R\sim 4/3\pi$ & \cite{FluidsBook} & Disc in $3$D fluid
\end{tabular}
\caption{{The drag coefficient ($\Lambda_R$) on a small particle ($R_p/R_m\ll 1$) as the Saffman length $\ell_{SD}$ is varied shows five different asymptotic behaviours. Investigation of the limits $\ell_{SD}\gg R_m^3/R_p^3$, $\ell_{Sd}\sim R_m^3/R_p^3$, and $\ell_{SD}=\mathcal O(R_p)$ represents an extension of the results in Ref.~\cite{levine2004mobility} for a point-like particle ($R_p=0$).}}\label{Drag coefficient asymptotics table}
\end{table}
\end{center}

\paragraph{Solvent-dominated limit {(5)}.}

{Finally, when the membrane viscosity is {very low} ($\varepsilon\gg 1$ or $\ell_{SD}\ll R_p$), we observe one more  {regime} (Fig.~\ref{Fig: Asymptotics}A, region {(5)}) where the membrane virtually disappears and $\Lambda_R$ is the same as the three-dimensional mobility of a disc in the solvent, namely $\Lambda_R\sim 4/3\pi$~\cite{FluidsBook}.}

{A summary of the identified asymptotic limits is provided in Table \ref{Drag coefficient asymptotics table}}. {An implication of our analysis is that the planar Saffman theory therefore captures the mobility of small inclusions as long as $\ell_{SD}$ is not too large, specifically $\ell_{SD}\ll R_m^3/R_p^2$.} We {also} note that the asymptotic behaviour in region {(5)}, corresponding to $\varepsilon\gg 1$ or $\ell_{SD}\ll R_p$ is a reflection of the  {spherical} geometry of our setup. Indeed, while it is evident from Eq.~(\ref{Stress Balance}) that the membrane applies negligibly small  {in-plane} stresses on the solvents, in-plane membrane incompressibility in general poses a non-trivial constraint on the ambient flows due to the no-slip condition. It is for instance well-known that a flat membrane with asymptotically small viscosity affects the leading-order translational resistance of a particle~\cite{CylinderInMembrane}. In a spherical geometry, on the other hand, any purely azimuthal surface flow is automatically divergence-free, so membrane incompressibility does not constrain solvent flow and the limit $\eta\to 0$ is regular (i.e.~the membrane simply disappears).

 {The membrane flow is qualitatively different in regions {(1)}-{(5)}. As shown in Fig.~\ref{Fig: Asymptotics}B, for values of $\varepsilon$ in region {(1)}, the membrane essentially rotates rigidly to reduce in-plane shear. This effect lingers in region {(2)}, but the increased shear noticeably slows down rotation. In region {(3)}, corresponding to $\theta_p^3\ll\varepsilon\ll 1$, the membrane transitions from the high curvature regime, $R_m\ll \ell_{SD}$ ($\varepsilon\ll \theta_p$), to the low curvature regime, $R_m\gg\ell_{SD}$ ($\varepsilon\gg \theta_p$). Interestingly, this transition is associated with a loss of monotonicity and the creation of a internal minimum near the particle inclusion~\cite{samanta2021vortex}, while rotation near the equator is nearly rigid to reduce the otherwise large in-plane shear. Finally, as the membrane viscosity is reduced further, the membrane flow decreases monotonically on the length scale of the particle with minimal qualitative differences between regions {(4)} and {(5)}.}

\subsubsection{ {Large particle limit: $R_p\sim R_m$}}\label{Low curvature limit}
We now consider the case of particle of size comparable to that of the membrane,  {similar to experimental observations of liquid domains in phase-separated giant unilamellar vesicles~\cite{cicuta2007diffusion}}. As already shown in Fig.~\ref{Fig:Drag Coefficient Many Values} and in \S~\ref{Small particle Limit}, when the membrane is very viscous ($\ell_{SD}/R_p\sim \ell_{SD}/R_m\gg 1$), the system rotates rigidly and $\Lambda_R\sim \theta_p^{-3}$. In this section, we investigate the opposite limit of a nearly inviscid membrane ($\ell_{SD}/R_p\sim \ell_{SD}/R_m\ll 1$). As explained in \S~\ref{Small particle Limit}, the membrane effectively disappears for small $\ell_{SD}$. It is therefore natural to compare our numerical results for $\Lambda_R$ with the  {exact} rotational mobility of a spherical cap $\hat\Lambda_R$ obtained in the absence of a membrane~\cite{SphericalCapTorque} 
\begin{equation}
\hat\Lambda_R(\theta_p)= \frac{1}{8\pi \theta_p^3}\left(8\theta_p-4\sin2\theta_p+\frac{16}{3}\sin^3\theta_p\right)\label{Collins Lambda} .
\end{equation}

{Consistently with the aforementioned result for a rotating disc in an unbounded fluid~\cite{FluidsBook}, we have $\hat\Lambda_R\to 4/3\pi$  as $\theta_p\to 0$, while $\hat\Lambda_R\to\theta_p^{-3}$ when $\theta_p\to\pi$ (i.e.~the particle becomes a spherical shell). We compare the  limit in Eq.~\eqref{Collins Lambda} with our numerical results in  Fig.~\ref{Fig: Asymptotics}C.
We see that very good agreement is obtained throughout the domain, confirming  that the membrane indeed vanishes as $\eta\to 0$.}

\section{Conclusion}\label{Discussion}

In this paper,  we computed the rotational mobility of a rigid particle embedded inside a spherical membrane for various particle sizes and Saffman-Delbr{\"u}ck  length  scales.  {The calculation was motivated by a number of relevant biological situations, such as the  movement of ATP synthase~\cite{RotatingATP}, aquaporin channels ~\cite{OptimizingPermeability, WaterPermeation, AquaporinsStructure},  {and} the Brownian motion of membrane-embedded particles~\cite{oppenheimer2009correlated}}. We started with the most general force-balance equations for the membrane and eventually obtained   an ODE for the membrane flow {$\mathbf v$}. After expanding the ambient {and membrane} flows with respect to a polynomial basis, we reduced the problem to an infinite set of linear  equations depending only on two dimensionless parameters: $\varepsilon$ and $\theta_p$.  {These correspond to $\varepsilon=2\mu R_p/\eta=2R_p/\ell_{SD}$, the ratio of the curvilinear particle radius and the Saffman-Debr{\"u}ck length, and $\theta_p=R_p/R_m$, the half-angle of the particle}. Using a  truncation  to a finite number of modes,  the resulting system could be   solved   {semi-analytically}, allowing us to  compute the dimensionless rotational mobility $\Lambda_R$ for many values of $\varepsilon$ and $\theta_p$,  {with results  summarised in Fig.~\ref{Fig:Drag Coefficient Many Values}}.  {We then explored the physical significance of the three relevant length scales -- curvilinear particle radius $R_p$, Saffman-Debr{\"u}ck length $\ell_{SD}$, and   membrane radius $R_m$ --   
     by considering different asymptotic limits. We demonstrated that the particle only sees the  spherical geometry of the membrane when $R_p=\mathcal O(R_m)$ or when the membrane viscosity is sufficiently high, while for sufficiently small $R_p\ll R_m$ the flow is purely local and the mobility is set by the planar limit~\cite{CylinderInMembrane}. The Saffman-Delbr{\"u}ck length operates as a cut-off length beyond which the flow is affected by the solvent traction. In particular, for small $\ell_{SD}$ the membrane disappeared altogether, while for large $\ell_{SD}$ the membrane rotated almost rigidly as if the solvents were not present}.

 {From a theoretical standpoint, this work sheds light on the interplay between particle geometry and intrinsic length scales in determining   local flow and particle mobility. This is made possible by considering a finite-sized particle, rather than a {point-like inclusion}~\cite{samanta2021vortex}. Computationally, we were able to numerically recover past results  for planar mobility and the torque on a spherical cap as special cases of our geometry~\cite{CylinderInMembrane, SaffmanDetailedCalculations, SphericalCapTorque}. The numerical implementation of a spherical membrane of vanishing viscosity (a limit which we showed to be regular unlike in the planar case~\cite{CylinderInMembrane}) may prove a valuable computational tool when dealing with rotational flow with spherical geometries.}

 Biologically, our work provides a way of estimating the viscosity of membranes containing large inclusions~\cite{cicuta2007diffusion}, which are affected by the membrane's geometry. Current experimental measurements of $\eta$ typically approximate the membrane as 
planar~\cite{LipidBilayer} and rely on Saffman's theory to estimate $\eta$ from the translational diffusion coefficient~\cite{SaffmanFlow, SaffmanDetailedCalculations, CylinderInMembrane} of membrane-bound {proteins and microspheres}~\cite{morris2015mobility}. When the inclusion is too large for Saffman's theory to apply, such as {for large proteins~\cite{shi2024drag, shi2022hydrodynamics}}, our work suggests that one may instead measure the rotational diffusion coefficient $D_R$ of the inclusion and exploit Einstein's relation $D_R=k_BT/G_p$~\cite{SaffmanFlow} to determine the torque, $G_p=k_BT/D_R$. Our predictions in Fig.~\ref{Fig:Drag Coefficient Many Values} may then be used to estimate $\varepsilon$, and thus $\eta$. {It should however be noted that measuring the rotational mobility may pose a significant challenge, requiring the invention of bespoke experimental protocols.}

An important extension of this work will consist of studying the other relevant mobility component, namely the translational mobility of the particle. This is expected to be significantly more challenging, as the resulting physical system will no longer be rotationally symmetric.  {It may likewise be physically relevant to analyse the case of {a non-circular particle~\cite{morris2015mobility}, or} a finite-sized inclusion protruding from the membrane, which may better model ATP synthase~\cite{RotatingATP} or membrane-embedded particles~\cite{oppenheimer2009correlated} than the case of a perfect embedding (Fig.~\ref{Fig:particle Schematic}). This is however expected to be rather challenging due to the more complex geometry.} Furthermore, one could allow the interior and exterior solvents to have different viscosities, as may be the case for cells whose cytoplasm  viscosity is larger than that in the surroundings.
Other directions for future work include the interactions of multiple rotating particles, {or adapting the theory to the case of non-rigid inclusions, such as liquid domains}~\cite{cicuta2007diffusion}. {The rotational diffusion of liquid domains arises as a result of random molecular torques \cite{SaffmanFlow}. Such torques are applied on extremely fast timescales, giving rise to oscillatory flows within the domain and the membrane.} An unsteady version of our theory will therefore be needed to capture this effect~\cite{kim2013microhydrodynamics, happel2012low}}. It may also be biologically relevant~\cite{powers2010dynamics} to allow for a more realistic membrane rheology, including viscoelastic effects~\cite{rey2006polar, rahimi2013curved}.\\

\funding{This work was funded in part by EPSRC (scholarship to MV).}

\appendix

\section{{Force and torque balance equations in the membrane}}\label{Force balance derivation}

  \begin{center}
 \begin{figure}[t]
 \hspace{1.5cm}
\includegraphics[scale=0.35]{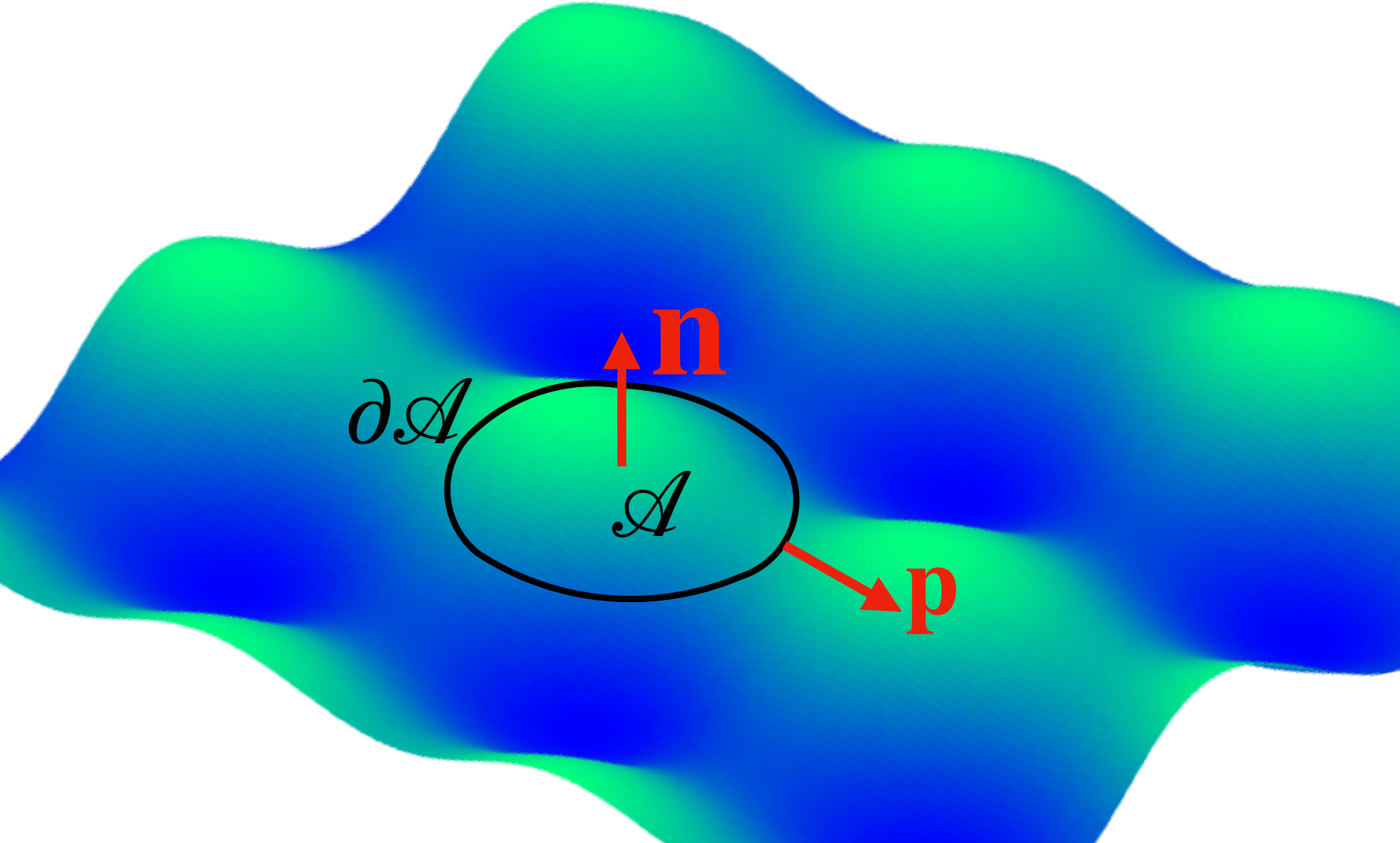}
\vspace{-0.2cm}
\caption{Force balance sketch: the membrane viscous forces exerted on the boundary of the area patch $\mathcal A$  {with unit normal $\mathbf n$, boundary $\p A$ and boundary unit normal $\mathbf p$} should balance the viscous forces generated by the solvents.  {Exploiting the arbitrariness of $\mathcal A$ yields the membrane Stokes equations (\ref{Stress Balance}).}}
\label{Fig:Force Balance Diagram}
\end{figure}
\end{center}
{For the purposes of obtaining a general theory, let the membrane be parametrised by arbitrary curvilinear coordinates $(x^1,x^2)\in \Sigma\subseteq \mathbb R^2$. Taking $\mathbf r$ to be the position vector in $\mathbb R^3$, we may endow the membrane with a coordinate basis $\mathbf e_a=\mathbf r_{,a}$. We further denote the unit normal to the membrane by $\mathbf n$, the metric tensor by $g_{ab}=\mathbf e_a\cdot\mathbf e_b$ and the extrinsic curvature tensor by $K_{ab}=\mathbf n\cdot\p_b\mathbf e_a$.} {The viscous stresses in the membrane are described by a stress tensor {$\sigma^{ab}$}, such that the force $\mathbf f$ per unit length on a curve with unit normal $\mathbf p$ (tangent to the membrane) is $\mathbf f=\sigma^{ab}p_a\mathbf e_b$}.
{Apply the force-balance condition to a small area patch $\mathcal A$ with boundary $\p \mathcal A$ and local boundary unit normal $p_a$ (Fig.~\ref{Fig:Force Balance Diagram})}. The external viscous force {on this patch} is {$\mathbf T\lVert \mathcal A\rVert+\mathcal O(\lVert \mathcal A\rVert^2)$}, where the norm denotes the area.  {By the standard divergence theorem in $\mathbb R^2$, letting $g=\det(g_{ab})$}, the viscous force due to stresses in the membrane is
\begin{align}
\oint_{\p \mathcal A}\sigma^{ab}\bold e_b p_a\mathrm ds&=\int_{{\Sigma}}\p_a(\sigma^{ab}\bold e_b \sqrt g)\mathrm dx^1\mathrm dx^2\label{Force on Boundary Intefral}\\
&=\int_{\mathcal A}\frac{1}{\sqrt g}\p_a(\sigma^{ab}\bold e_b \sqrt g)\mathrm dS=\frac{1}{\sqrt g}\p_a(\sigma^{ab}\bold e_b \sqrt g)\lVert \mathcal A\rVert+\mathcal O(\lVert \mathcal A\rVert^2).
\end{align}
 {Assuming the membrane is viscous enough for inertia to be negligible, forces must locally balance.  {As a result, for} $\lVert \mathcal A\rVert\to 0$,
\begin{align}
&\frac{1}{\sqrt g}\p_a(\sigma^{ab}\bold e_b \sqrt g)+ {\mathbf T}=\mathbf 0\label{Force balance simpler}\\
\Rightarrow\ & (\p_a\sigma^{ab}+\Gamma^c_{ac}\sigma^{ab}+\Gamma^b_{ac}\sigma^{ac})\mathbf e_b+\sigma^{ab}K_{ab}\mathbf n+ {\mathbf T}=\mathbf 0\label{derivative of sqrt g step}\\
\Rightarrow\ & (\nabla_a\sigma^{ab})\mathbf e_b+\sigma^{ab}K_{ab}\mathbf n+ {\mathbf T}=\mathbf 0 \label{Force Balance Proof}
, \end{align}}
{where we exploited the fact that $\p_a\sqrt g=\Gamma^b_{ab}\sqrt g$ in Eq.~\eqref{derivative of sqrt g step}. The expression in Eq.~(\ref{Force Balance Proof}) is now equal to Eq.~(\ref{Stress Balance}) in the main body of the paper.}

{An equivalent way to express force balance, more reminiscent of the standard Stokes equations, is obtained by rewriting Eq.~(\ref{Force balance simpler}) as
\begin{align}
 {\text{div}_s}(\sigma^{ab}\mathbf e_a\otimes\bold e_b)+ {\mathbf T}=\mathbf 0\label{Local Force Balance},  
 \end{align}
where the operator $ {\text{div}_s}=\mathbf e^a\p_a$ denotes the surface derivative~\cite{al2023morphodynamics} (not to be confused with the covariant derivative). To see why Eq.~(\ref{Local Force Balance}) holds, note that from the Leibniz rule for partial derivatives we have
 \begin{align}
 {\text{div}_s}(\sigma^{ab}\mathbf e_a\otimes\bold e_b)=(\nabla_c\sigma^{ab})\mathbf e^c\otimes\mathbf e_a\otimes\mathbf e_b+\sigma^{ab}\mathbf e^c\otimes (K_{ab}\mathbf n)\otimes\mathbf e_b+\sigma^{ab}\mathbf e^c\otimes \mathbf e_a\otimes (K_{bc}\mathbf n)  ,
 \end{align}
 and therefore, contracting the first tensor product
  \begin{align}
 {\text{div}_s}(\sigma^{ab}\mathbf e_a\otimes\bold e_b)=(\nabla_a\sigma^{ab})\mathbf e_b+\sigma^{ab}K_{ab}\mathbf n .
 \end{align}
 }
 {Finally,} integrating Eq.~(\ref{Local Force Balance}) over the membrane, we obtain that the total external force $\int_{\mathcal M}{\mathbf T}\mathrm dS$ reduces to a boundary term, and hence vanishes for a closed membrane with no {inclusions}.

{By a similar argument as for the forces}, torques balance within the membrane, as for any area patch $\mathcal B$
\begin{align}
 \int_{\p\mathcal B}\bold r\times \sigma^{ab}\bold e_{b}p_a\mathrm ds&=\int_{\mathcal B}\partial_{a}(\bold r\times \sigma^{ab}\bold e_{b}\sqrt{g})\mathrm dx^1\mathrm dx^2\nonumber\\
 &=\int_{\mathcal B} \sigma^{ab}\bold e_a\times\bold e_b\sqrt{g}\mathrm dx^1\mathrm dx^2+\int_{\mathcal B}\bold r\times \frac{1}{\sqrt g}(\sigma^{ab}\bold e_b\sqrt g)\mathrm dS\label{r,a step}\\
 &=\int_{\mathcal B}\bold r\times   {\text{div}_s}(\sigma^{ab}\mathbf e_a\otimes\bold e_b)\mathrm dS\nonumber\\
 &=\int_{\mathcal B}-\bold r\times {\mathbf T}\mathrm dS.
\end{align}
Note that in Eq.~(\ref{r,a step}) we have used the identity $\bold r_{,a}=\bold e_a$ and the first integral cancels due to symmetry of $\sigma^{ab}$. {For our problem, this means that the total torque on the particle is equal to the torque on the membrane-particle system from the external solvent.}

\section{{Membrane flow equation as limit of the $3$D Stokes equations}}\label{Membrane Flow Equation as Limit of the $3$D Stokes Equations}
 {To further elucidate the two-dimensional flow model for the membrane, in this Appendix we derive Eq.~(\ref{NS phi direction}) as the limit of the standard three-dimensional Stokes equations.} 
 
  {We start by approximating the membrane as a thin layer of Newtonian fluid in a spherical annulus {$R_m\leq r\leq R_m+\Delta R_m$} with viscosity $\overline{\eta}$. Because of the presence of undeformable molecules oriented normal to the membrane~\cite{LipidBilayer}, we assume that such a fluid layer must have vanishing radial shear; using $\mathbf E$ to denote the 
 rate-of-strain tensor, this means  that $\mathbf E_{rr}=\mathbf E_{r\theta}=\mathbf E_{r\phi}=0$. In a rotationally symmetric setup, this constrains the flow to be of the form
\begin{align}
{\bold u}&{=rv(\theta)\sin\theta\mathbf e_{\phi}}, & {R_m\leq}&{r\leq R_m+\Delta R_m,}
\end{align}
 Flow in the membrane is affected by the traction from the solvents, which can be thought of as a force per unit volume $f\boldsymbol{\phi}$. The corresponding forced three-dimensional membrane Stokes equations {$\nabla\overline p-\overline{\eta}\nabla^2\mathbf u=f\mathbf e_{\phi}$} is now
\begin{align}
{\overline{\eta}\left(v''\sin\theta+3v'\cos\theta\right)+rf=0.}
\end{align}
In the limit of a thin membrane  {with $\Delta R_m/R_m\ll 1$}, the total force on a volume element should be the same as the traction on the upper and lower surfaces, i.e.
\begin{align}
{f\Delta R_m=\tau^{\phi}}
\end{align}
Assuming a very viscous membrane, sending now $\Delta R_m\to 0$, so that  {$r\to R_m$}, while keeping the effective two-dimensional membrane viscosity   {$\eta=\overline{\eta} \Delta R_m$} constant leads to 
\begin{align}
{\eta(v''\sin\theta+3v'\cos\theta)+R_m\tau^{\phi}=0.}
\end{align}

This coincides with Eq.~(\ref{NS phi direction}) in the main text, showing that the previously introduced two-dimensional model of the membrane is obtained as the limit of a thin, non-shearing three-dimensional flow.

\section{{Derivation of governing Eqs.~(\ref{c_n Recursion}) and (\ref{Navier Stokes Expanded Forcing1})}}\label{Legendre Polynomial Expansion}

In order to determine the flow in the membrane and the solvents, we need to find the expansion coefficients $c_n$. These can be expressed in terms of $v(x)$ by mutliplying Eq.~(\ref{Expansion to Test Against}) by $1-x^2$ and taking the inner product with $P_m$. Using the classical equalities~\cite{riley2006mathematical, arfken1985mathematical}
\begin{align}
&(1-x^2)P_n'=n(P_{n-1}-xP_n),\\
&\int_{-1}^1P_n(x)P_m(x)\mathrm dx=\frac{2\delta_{n,m}}{2n+1},\label{Orthogonality condition 1}\\
&\int_{-1}^{1}xP_n(x)P_m(x)\mathrm dx=\frac{2(m+1)}{(2m+1)(2m+3)}\delta_{n,m+1}+\frac{2m}{(2m-1)(2m+1)}\delta_{n,m-1}\label{Orthogonality condition 2},
\end{align}
for $n,m\geq 0$, we readily obtain the recursive relationship  for $n\geq 0$
\begin{align}
\frac{(n+1)(n+2)}{(2n+1)(2n+3)}c_{n+1}-\frac{n(n-1)}{(2n-1)(2n+1)}c_{n-1}=\frac{1}{2}\int_{-1}^1P_n(x)v(x)(1-x^2)\mathrm dx  \label{c_n Recursion appendix}.
\end{align}
The membrane velocity $v(x)$ and the  {coefficients} $c_n$ are also coupled via the momentum  {equation}, Eq.~\eqref{NS phi direction}. Substituting $x=\cos\theta$, this takes the form
\begin{align}
{\eta[(1-x^2)v''(x)-4xv'(x)]+ {R_m}(1-x^2)^{-1/2}\tau^{\phi}}&{=0} &\theta_p<\theta\leq\pi {\text{ (membrane)}}\label{NS phi direction appendix}.
\end{align}
The azimuthal tractions may be evaluated as
\begin{align}
{(1-x^2)^{-1/2}\tau^{\phi}}&{=\mu R_m\left[\frac{\partial(V^{\phi}_+-V^{\phi}_-) }{\partial r}\right]_{r=R_m}=-\mu\sum_{n=1}^{\infty}(2n+1)c_nP_n'(x)}\label{Sum of Tractions Expansion appendix},
\end{align}
finally yielding
\begin{align}
(1-x^2)v''(x)-4xv'(x)=\frac{\mu R_m}{\eta}\sum_{n=1}^{\infty}(2n+1)c_nP_n'(x)&=0 &x\geq x_p {\text{ (membrane)}}\label{NS phi direction appendix},
\end{align}
which is the same as Eq.~(\ref{Navier Stokes Expanded Forcing1}) in the main text.

\bibliographystyle{RS}
\bibliography{membrane_paper_references.bib}

\enlargethispage{20pt}

\end{document}